\documentclass[preprint,proceedings]{rmaa}

\usepackage{paralist}

\usepackage{psfrag,color}

\SetYear{2004}
\SetConfTitle{Compact Binaries in the Galaxy and Beyond}

\title{Results of a search for new microquasars in the Galaxy} 

\author{
  M. Rib\'o,\altaffilmark{1}
  J. M. Paredes,\altaffilmark{2}
  J. Mart\'{\i},\altaffilmark{3}
  J. Casares,\altaffilmark{4}
  J. S. Bloom,\altaffilmark{5,6}
  E. E. Falco,\altaffilmark{7}
  E. Ros,\altaffilmark{8}
  and M. Massi,\altaffilmark{8}}

\altaffiltext{1}{Service d'Astrophysique, CEA Saclay, B\^at. 709, L'Orme des
Merisiers, 91191 Gif-sur-Yvette, Cedex, France.}
\altaffiltext{2}{Universitat de Barcelona, Barcelona, Spain.}
\altaffiltext{3}{Universidad de Ja\'en, Ja\'en, Spain.}
\altaffiltext{4}{Instituto de Astrof\'{\i}sica de Canarias, La Laguna, Spain.} 
\altaffiltext{5}{Harvard Society of Fellows, Cambridge, USA.}
\altaffiltext{6}{Harvard-Smithsonian CfA, Cambridge, USA.}
\altaffiltext{7}{F. L. Whipple Observatory, Amado, USA.}
\altaffiltext{8}{MPIfR, Bonn, Germany.}

\shortauthor{Rib\'o et~al.}
\shorttitle{Results of a search for new microquasars in the Galaxy}

\listofauthors{M. Rib\'o, J. M. Paredes, J. Mart\'{\i}, J. Casares, J. S. Bloom, E. E. Falco, E. Ros, \& M. Massi}
\indexauthor{Rib\'o, M.}
\indexauthor{Paredes, J. M.}
\indexauthor{Mart\'{\i}, J.}
\indexauthor{Casares, J.}
\indexauthor{Bloom, J. S.}
\indexauthor{Falco, E. E.}
\indexauthor{Ros, E.}
\indexauthor{Massi, M.}

\abstract{
We present here the results of a search for new microquasars at low galactic
latitudes, based on a cross-identification between the {\it ROSAT} all sky
Bright Source Catalog (RBSC) and the NRAO VLA Sky Survey (NVSS) and follow-up
observations. The results obtained up to now suggest that persistent/silent
microquasars such as LS~5039 are rare objects in our Galaxy, and indicate that
future deeper surveys, and harder than the RBSC in X-rays, will play a
fundamental role in order to discover them.
}

\addkeyword{Radio continuum: stars}
\addkeyword{X-rays: binaries}

\suppressfulladdresses

\begin{document}
\maketitle

\section{Introduction} \label{sec:intro}

Microquasars are X-ray binary systems with the ability to generate
relativistic jets (see Mirabel \& Rodr\'{\i}guez 1999 and Fender 2004 for
detailed reviews on the topic). These sources mimic, on smaller scales, many
of the phenomena seen in AGNs and quasars, but with timescales several orders
of magnitude shorter (since $t$ scales with $M$). This property allows us to
study, in a few minutes, the accretion/ejection processes that take place near
galactic compact objects. Unfortunately, the population of known microquasars
is still very small with merely around 15 cases. Therefore, it is worth to
search for new microquasars in order to increase the known population to allow
meaningful statistical studies.

\section{Cross-identification method} \label{sec:cross}

In order to identify new microquasar candidates we have to find new radio
emitting X-ray binaries. To this end, we have performed a cross-identification
between the X-ray catalog RBSC (Voges et~al.\@ 1999) and the radio catalog
NVSS (Condon et~al.\@ 1998), which covers the sky north of
$\delta=-40^{\circ}$. We have adopted the following selection criteria:
1)~Absolute galactic latitude $<5^{\circ}$; 2)~No screening flags about nearby
sources contaminating measurements or problems with position determinations in
the RBSC; 3)~Hardness ratios, $HR1+\sigma(HR1)$, higher than 0.9; 4)~NVSS
sources within the 2$\sigma$ error boxes of the RBSC sources; 5)~Unresolved by
the NVSS. The sources selected with the RBSC/NVSS cross-identification, a
total of 35, were then filtered with complementary optical information using
the following criteria: 1)~No extragalactic information within SIMBAD and NED;
2)~No extended appearance in the Digitized Sky Survey, DSS1 and DSS2-red
images. Only 17 sources fulfilled all the selection criteria, including the
well known microquasars LS~5039, SS~433 and Cygnus~X-3, and the new
microquasar LS~I~+61~303. Therefore, our method recovered all persistent radio
emitting high mass X-ray binaries except Cygnus~X-1, which was too faint
during the NVSS observations.

\begin{table*}[t!]\centering
\setlength{\tabnotewidth}{\textwidth}
\tablecols{9}
\caption{Summary of the obtained results\tabnotemark{a}}
\label{table:summary}
\begin{tabular}{ll@{~}c@{~~~}l@{~~}cl@{~~}c@{~~}cl}
\toprule
1RXS name           & \multicolumn{2}{c}{VLA obs.}    & \multicolumn{2}{c}{Optical obs.}  & \multicolumn{3}{c}{EVN+MERLIN obs.}                  & Opt. spec. \\
                    & Structure         & $\alpha$    & Structure    & $I$                & Structure            & $\beta$ & $\theta[^{\circ}]$  & \\
\midrule
J001442.2+580201    & compact           & $-0.2$      & point-like   & 19.9               & 2-sided jet          & $>0.20$ & $<78$               & Featureless \\
J013106.4+612035    & compact           & $-0.1$      & point-like   & 17.9               & 1-sided jet          & $>0.31$ & $<72$               & Featureless \\
J042201.0+485610    & compact           & ~\,$+1.5^*$ & extended$^*$ & 17.5               & not detected$^*$     & \multicolumn{1}{c}{\nodata} & \multicolumn{1}{c}{\nodata} & Seyfert 1 \\
J062148.1+174736    & compact           & $+0.1$      & extended$^*$ & 17.6               & compact              & \multicolumn{1}{c}{\nodata} & \multicolumn{1}{c}{\nodata} & Featureless \\
J072259.5$-$073131  & 1-sided jet$^*$   & $-0.2$      & point-like   & 16.8               & bent 1-sided jet$^*$ & $>0.29$ & $<73$               & BL Lac (?)\\
J072418.3$-$071508  & compact           & $+0.1$      & point-like   & 17.2               & bent 1-sided jet$^*$ & $>0.51$ & $<60$               & FSRQ \\
\bottomrule
\tabnotetext{a}{An asterisk indicates a non-expected behavior for microquasars.}
\end{tabular}
\end{table*}

\section{Follow-up observations} \label{sec:obs}

Follow up VLA~A configuration multifrequency and multiepoch observations of 6
of the remaining 13 unknown sources, performed in July 1999, provided accurate
positions, as well as spectral and variability information. This allowed us to
discover optical counterparts for all of them with probabilities of random
coincidence below 0.3\% after our own optical observations conducted with the
1.5\,m OAN (November 1998) and 2.2\,m CAHA (December 1999) telescopes at Calar
Alto (Paredes et~al.\@ 2002).

With the aim of revealing possible jet-like features at milliarcsecond scales,
we observed these six sources with the EVN and MERLIN on 2000 February
29th/March 1st at 5~GHz. We detected five of the six observed sources, which
showed different morphologies: one source with a two-sided jet, three sources
having a one-sided jet and a compact source (Rib\'o et~al.\@ 2002).

Finally, we performed optical spectroscopic observations of these six sources
with the 4.2\,m WHT and 6.5\,m MMT telescopes (Mart\'{\i} et~al.\@ 2004).

\section{Summary and conclusions} \label{sec:summary}

After a detailed analysis of all the follow-up observations, we show in
Table~\ref{table:summary} a summary of the obtained results. The first two
sources, 1RXS J001442.2+580201 and 1RXS J013106.4+612035, show featureless
spectra indicative of an extragalactic nature. Nevertheless, the possibility
of having highly reddened stars cannot be completely excluded (see Mart\'{\i}
et~al.\@ 2004 for details). 1RXS J042201.0+485610 is a Seyfert~1 galaxy with
broad Hydrogen emission lines and $z=0.114$. The source 1RXS J062148.1+174736
is probably an extragalactic object due to the featureless spectrum and
extended nature of the optical counterpart. 1RXS J072259.5$-$073131 shows
properties common to BL Lac objects, while 1RXS J072418.3$-$071508 is an
already identified Flat Spectrum Radio Quasar (FSRQ) with $z=0.270$.


Assuming that none of our candidates is galactic, it appears that the
population of new and persistent microquasars is not very numerous in the
Galaxy. The corresponding density of new (bright) Cygnus~X-3 and (faint)
LS~3039-like system is constrained to be $\la 1.1 \times 10^{-12}$~pc$^{-3}$
and $\la 5.6 \times 10^{-11}$~pc$^{-3}$, respectively. Although we plan to
expand our cross-identification studies to $5^{\circ} \leq |b| \leq
10^{\circ}$, the basic limitation of the RBSC low-energy range, where X-ray
photons are highly absorbed, will persist. Therefore, sensitive surveys in
hard X-rays and $\gamma$-rays, such as the current {\it INTEGRAL} Galactic
Plane Survey or the planned {\it EXIST} mission, will play a fundamental role
in order to reveal the {\it real} population of persistent microquasars in the
Galaxy.

\acknowledgements

M.~R., J.~M.~P. and J.~M. acknowledge partial support by DGI of the Ministerio
de Ciencia y Tecnolog\'{\i}a (Spain) under grant AYA2001-3092, as well as
partial support by the European Regional Development Fund (ERDF/FEDER).
M.~R. acknowledges support by a Marie Curie Fellowship of the European Community programme Improving Human Potential under contract number HPMF-CT-2002-02053.
J.~M. is supported by the Junta de Andaluc\'{\i}a (Spain) under project
FQM322, and has been also aided in this work by an Henri Chr\'etien
International Research Grant administered by the AAS.

\end{document}